\documentclass[aps,prl,twocolumn,superscriptaddress]{revtex4-1}
\usepackage{graphicx}
\usepackage[hidelinks]{hyperref} 
\usepackage{bm}
\usepackage{amsmath}
\graphicspath{{./}}

\begin{document}


\title{Ordering phenomena of spin trimers accompanied by large geometrical Hall effect}

\renewcommand*{\thefootnote}{\arabic{footnote}}

\author{Shang Gao}
\thanks{These authors contributed equally to this work.\\ 
Corresponding authors (emails): shang.gao@riken.jp, and maximilian.hirschberger@riken.jp }
\affiliation{RIKEN Center for Emergent Matter Science, Wako 351-0198, Japan}

\author{Max Hirschberger}
\thanks{These authors contributed equally to this work.\\ 
Corresponding authors (emails): shang.gao@riken.jp, and maximilian.hirschberger@riken.jp }
\affiliation{RIKEN Center for Emergent Matter Science, Wako 351-0198, Japan}

\author{Oksana Zaharko}
\affiliation{Laboratory for Neutron Scattering and Imaging, Paul Scherrer Institut, CH-5232 Villigen PSI, Switzerland}

\author{Taro Nakajima}
\affiliation{RIKEN Center for Emergent Matter Science, Wako 351-0198, Japan}

\author{Takashi Kurumaji}
\thanks{Current address: Department of Physics, Massachusetts Institute of Technology, Cambridge, Massachusetts 02139, USA}
\affiliation{RIKEN Center for Emergent Matter Science, Wako 351-0198, Japan}

\author{Akiko Kikkawa}
\affiliation{RIKEN Center for Emergent Matter Science, Wako 351-0198, Japan}

\author{Junichi Shiogai}
\affiliation{Institute for Materials Research, Tohoku University, Sendai 980-8577, Japan.}

\author{Atsushi Tsukazaki}
\affiliation{Institute for Materials Research, Tohoku University, Sendai 980-8577, Japan.}

\author{Shojiro Kimura}
\affiliation{Institute for Materials Research, Tohoku University, Sendai 980-8577, Japan.}

\author{Satoshi Awaji}
\affiliation{Institute for Materials Research, Tohoku University, Sendai 980-8577, Japan.}

\author{Yasujiro Taguchi}
\affiliation{RIKEN Center for Emergent Matter Science, Wako 351-0198, Japan}

\author{Taka-hisa Arima}
\affiliation{RIKEN Center for Emergent Matter Science, Wako 351-0198, Japan}
\affiliation{Department of Advanced Materials Science, University of Tokyo, Kashiwa 277-8561, Japan}

\author{Yoshinori Tokura}
\affiliation{RIKEN Center for Emergent Matter Science, Wako 351-0198, Japan}
\affiliation{Department of Applied Physics and Tokyo College, University of Tokyo, Tokyo 113-8656, Japan}


\date{\today}

\begin{abstract}

The wavefuntion of conduction electrons moving in the background of a non-coplanar spin structure can gain a quantal phase -- Berry phase -- as if the electrons were moving in a strong fictitious magnetic field. Such an emergent magnetic field effect is approximately proportional to the solid angle subtended by the spin moments on three neighbouring spin sites, termed the scalar spin chirality. The entire spin chirality of the crystal, unless macroscopically canceled, causes the geometrical Hall effect of real-space Berry-phase origin, whereas the intrinsic anomalous Hall effect (AHE) in a conventional metallic ferromagnet is of the momentum-space Berry-phase origin induced by relativistic spin-orbit coupling (SOC).  Here, we report the ordering phenomena of the spin-trimer scalar spin chirality and the consequent large geometrical Hall effect in the breathing kagom\'e lattice compound Dy$_3$Ru$_4$Al$_{12}$, where the Dy$^{3+}$ moments form non-coplanar spin trimers with local spin chirality. Using neutron diffraction, we show that the local spin chirality of the spin trimers as well as its ferroic/antiferroic orders can be switched by an external magnetic field, accompanying  large changes in the geometrical Hall effect. Our finding reveals that systems composed of tunable spin trimers can be a fertile field to explore large emergent electromagnetic responses arising from real-space topological magnetic orders. 

\end{abstract}


\maketitle


Conventional electronic devices are based mainly upon the band dispersions of the conducting electrons and not on their phase factors. However, better understanding of the topological character of  electron bands that has been achieved in recent years is now promising next-generation devices where both the dispersions and the phase factors can be tailored \cite{qi_topological_2011, tokura_emergent_2017}. Such a prospect is derived from the fact that the Berry phase, which describes the change of the phase factor for an electron moving adiabatically around a loop in real or reciprocal space, has a direct impact on the electron transport \cite{xiao_berry_2010}. Specifically, in ferromagnetic systems with broken time-reversal symmetry, it has been established that a non-zero Berry phase over the occupied bands in reciprocal space can be induced by the relativistic SOC, which gives rise to an intrinsic AHE with the transverse conductivity $\sigma_{xy}$ proportional to the spin polarization of the conducting electron~ \cite{fang_anomalous_2003, yao_first_2004,nagaosa_anomalous_2010}.

Non-zero Berry phases and a consequent Hall effect can also arise in a special type of magnet where the magnetic moments form non-coplanar structures. In these magnets, the topologically non-trivial Berry phase is induced not by the SOC as in the case of the `conventional' AHE, but by non-zero scalar spin chirality $\chi_{ijk} = \bm{S}_i \cdot (\bm{S}_j\times \bm{S}_k)$, with $\bm{S}$ denoting the localized spins at vertices $i$, $j$, and $k$ of a triangle \cite{ye_berry_1999, ohgushi_spin_2000, shindou_orbital_2001,tatara_chirality_2002, onoda_anomalous_2004}. Hereafter we refer to such scalar spin chirality simply as chirality. This chirality-induced  Hall effect, here named geometrical Hall effect (GHE), is readily understood in the real space picture. Due to the coupling with the localized moments, an electron hopping successively across the triangular sites $i$-$j$-$k$-$i$ will gain a Berry phase that is approximately proportional to the solid angle spanned by the localized moments as if the electrons were circling around a magnetic flux \cite{ohgushi_spin_2000}.

\begin{figure}[t!]
\includegraphics[width=0.48\textwidth]{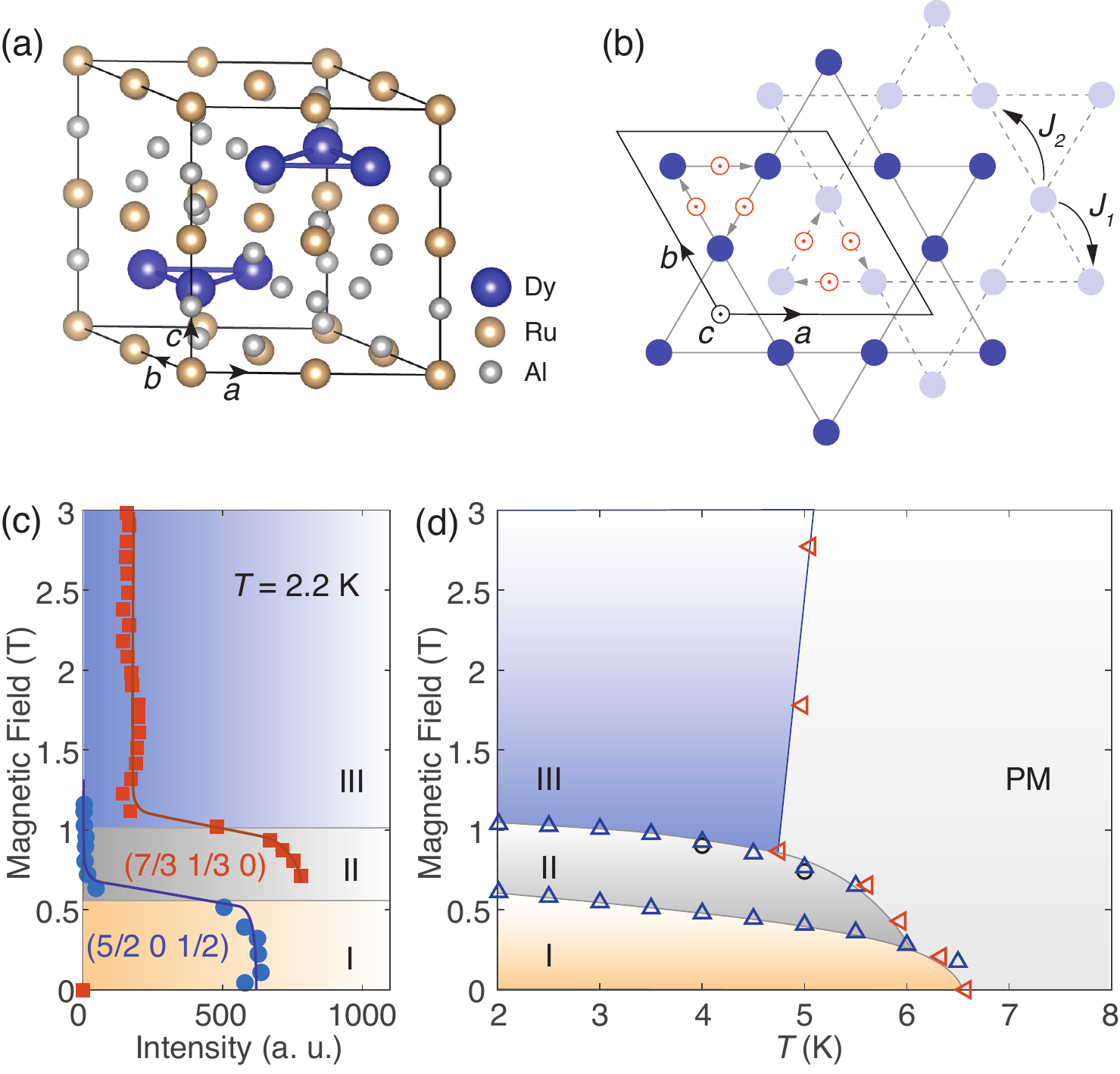}
\caption{(color online). (a) Crystal structure of Dy$_3$Ru$_4$Al$_{12}$ with space group $P6_3/mmc$. (b) Dy$^{3+}$ ions on the $6h$ Wyckoff sites form breathing kagom\'e layers in the $ab$ plane. Dy$^{3+}$ ions with $z = 0.25$ and $z = 0.75$ are shown by dark and light blue circles, respectively. In the unit cell enclosed with solid lines, directions of the Dzyaloshinskii-Moriya Interaction (DMI) vector defined over the nearest-neighbour bonds are shown in red circles, and the corresponding bond directions are shown by grey arrows over the bonds. The nearest-neighbour $J_1$ and second-neighbour $J_2$ bonds  are indicated by curved arrows. (c) Field-dependence of the neutron diffraction intensities of the magnetic reflections (5/2 0 1/2) and (7/3 1/3 0) measured at $T = 2.2$~K. (d) Phase diagram of Dy$_3$Ru$_4$Al$_{12}$ obtained from magnetization (up-pointing triangles) and heat capacity (left-pointing triangles) measurements~\cite{supp}. PM represents the paramagnetic phase. Effect of the demagnetization field has been corrected~\cite{supp}. Error bars representing standard deviations are smaller than the symbol size.
}
\label{fig:structure_phase}
\end{figure}

In spite of continuous efforts~\cite{matl_hall_1998, taguchi_spin_2001,grohol_spin_2005, takatsu_unconventional_2010, machida_time_2010, shiomi_hall_2012, ueland_controllable_2012, surgers_large_2014}, an unambiguous experimental illustration of the correspondence between the non-coplanar spin trimers  and the GHE is still missing. This is partly due to the difficulty in stabilizing commensurate non-coplanar magnetic structures on two-dimensional lattices \cite{grohol_spin_2005}. For example, on the prototypical kagom\'e lattice, Heisenberg spins normally prefer to order in a coplanar structure due to its amenability to fluctuations \cite{reimers_order_1993, elhajal_symmetry_2002, essafi_generic_2017}, and the coplanarity is maintained even in a magnetic field \cite{zhitomirsky_field_2002}.

In this letter, we show that non-coplanar spin trimers can be stabilized in the intermetallic compound Dy$_3$Ru$_4$Al$_{12}$, where the magnetic Dy$^{3+}$ ions constitute a breathing kagom\'e lattice with corner-sharing triangles of two different sizes (see Figs. \ref{fig:structure_phase}a and b)~\cite{gorbunov_electronic_2014, ishii_magnetic_2018}. 
Using neutron diffraction, we find that both the stacking order and the local chirality of the spin trimers can be tuned by a magnetic field. As long as the entire scalar spin chirality becomes non-zero, a large GHE emerges in our magneto-transport measurements. The geometrical origin of the observed Hall effect is also confirmed through a semi-quantitative comparison between the magnitude of the Hall conductivity and the entire scalar spin chirality.

Our Dy$_3$Ru$_4$Al$_{12}$ single crystals were grown using the Czochralsky technique~\cite{supp}. Neutron diffraction experiments on a single crystal sample of Dy$_3$Ru$_4$Al$_{12}$ were performed on the thermal-neutron diffractometer ZEBRA at the Swiss Spallation Neutron Source SINQ of the Paul Scherrer Institut PSI. Incoming neutron wavelength of 1.18 \AA\ (Ge(311) monochromator) was used for the measurements. Magneto-transport experiments were performed on single crystals with characteristic dimensions of 3.0$\times$0.8$\times$0.15 mm$^3$, where the largest faces were perpendicular to the crystallographic $c$-axis and were aligned perpendicular to the magnetic field. The electric current was applied along the $a^*$ axis in reciprocal space. Measurements below 14~T were performed on the Quantum Design PPMS, and measurements up to 24~T shown in the Supplemental Materials were performed at the High Field Laboratory for Superconducting Materials at Tohoku University~\cite{supp}.

At zero field, Dy$_3$Ru$_4$Al$_{12}$ is known to enter a magnetic long-range ordered state with a propagation vector of (1/2 0 1/2)~\cite{gorbunov_electronic_2014}. However, by applying a magnetic field along the $c$ axis, we found in the present study that the magnetic propagation vector can be shifted from $\bm{q}_1 =$ (1/2 0 1/2) to $\bm{q}_2=$ (1/3 1/3 0). As is shown in Fig.~\ref{fig:structure_phase}c, our neutron diffraction experiments reveal that at temperature $T$ = 2.2 K, the intensity of the (5/2 0 1/2) reflection in phase I drops to zero at a field of $\mu_0H\approx$ 0.6~T, while a new reflection emerges at (7/3 1/3 0). The intensity of the  (7/3 1/3 0) reflection is not constant, but decreases sharply at $\mu_0H\approx 1.2$ T while remaining finite, indicating the appearance of two distinct field-induced phases (II and III), consistent with the magnetic transitions observed by magnetization measurements~\cite{gorbunov_electronic_2014} (see also Fig.~\ref{fig:hall}a). Following the anomalies in the magnetic susceptibility and heat capacity across the phase transitions~\cite{supp}, we map out the $H$-$T$ phase diagram as presented in Fig.~\ref{fig:structure_phase}d.

Neutron diffraction datasets were collected in the three phases at $T= 2.2$ K to clarify their precise magnetic structures. Figure~\ref{fig:magstr} summarizes our refinement results. Details for the dataset refinement can be found in the Supplemental Materials~\cite{supp}. We notice that the magnetic structures in both phases I and II consist of similar spin trimers with non-zero local chirality. As is presented in Fig.~\ref{fig:magstr}d, four spin-trimer configurations are observed in the magnetic structure of phases I and II: the in-plane spin components $S_{ab}$ are all pointing inwards or outwards, and the out-of-plane components $S_c$ aligning uniformly parallel or anti-parallel to the $c$ axis. From our refinements, the ratio $|S_c/S_{ab}|$ remains nearly constant at $\sim 2$ in both phases I and II, meaning the magnitude $\chi_0$ of the local chirality does not change across the transition between phases I and II. Therefore, the entire chirality is completely determined by the stacking pattern of the trimers. In phase I with $\bm{q}_1 =$ (1/2 0 1/2), the sign of the local chirality is reversed between neighbouring unit cells along the $a$ and $c$ axes, implying full cancellation of the local chirality. However, in phase II with $\bm{q}_2 =$ (1/3 1/3 0), the signs are arranged in a sequence of $+$/$+$/$-$ along $a$ and $b$ axes, which results in an average chirality per spin-trimer unit of $\chi_0/3$ with $\chi_0 \approx 0.17$. Here the chirality $\chi_0$ is calculated assuming a unit spin length $|S|=1$. Over the larger second-neighbour triangles, the global chirality also becomes non-zero in phase II, and its magnitude equals $\chi_0/3$, as large as that over the spin-trimer unit.

\begin{figure*}[t!]
\includegraphics[width=0.85\textwidth]{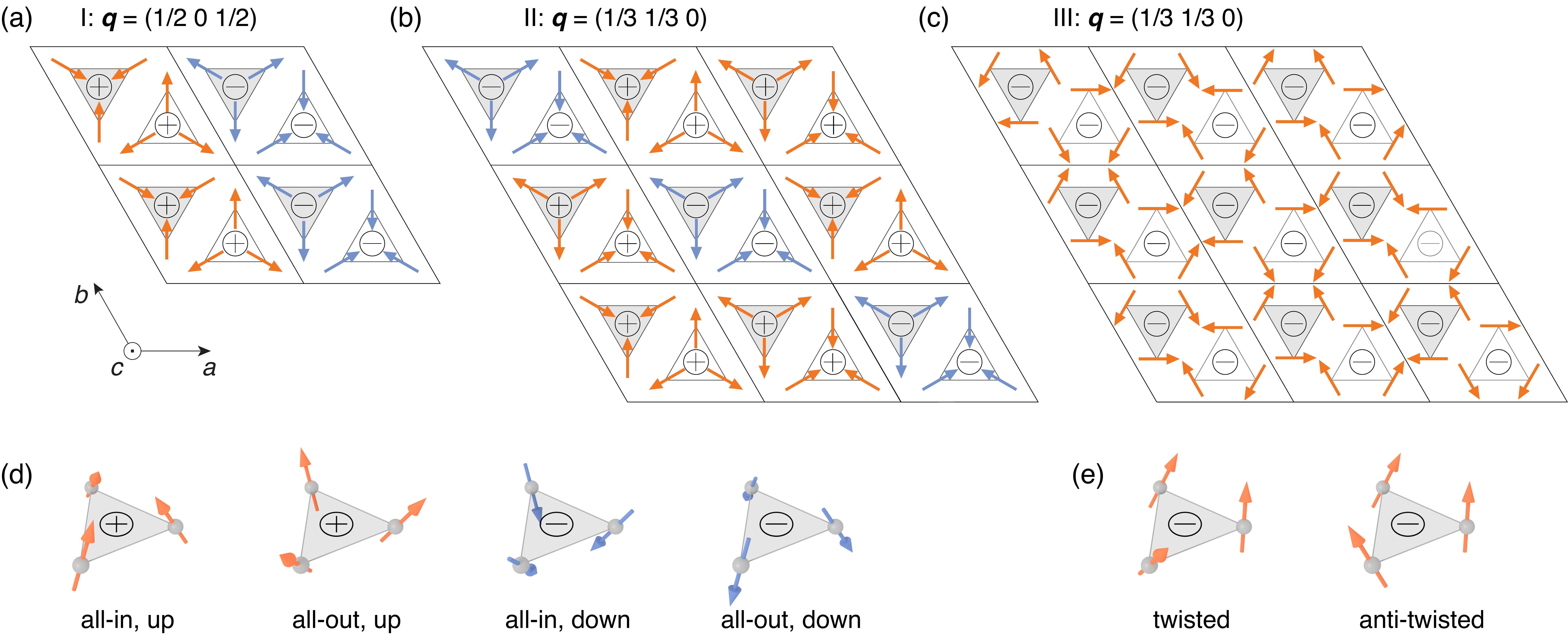}
\caption{(color online). (a) Magnetic structure of phase I with $\bm{q}_1 =$ (1/2 0 1/2). Shaded (blank) triangles indicate spin trimers within the $z =0.25$ ($z = 0.75$) layer. Spins with a positive (negative) component along the $c$ direction are shown by the orange (blue) arrows. Signs of the scalar spin chirality over the trimers are indicated at the center of the triangles. (b,c)  Magnetic structures of phases II (b) and III (c) with $\bm{q}_2 =$ (1/3 1/3 0). (d) Spin trimers in phases I and II. In these two phases, spin components in the $ab$ plane are either all pointing towards the center of the triangles (all-in) or out of the triangles (all-out).  (e) Twisted and anti-twisted spin trimers in phase III where the spin components in the $ab$ plane are parallel to the opposite edge of the triangles.
In phases I, II, and III, the  refined magnitude of the ordered Dy$^{3+}$ moments are 6.5(5), 9.3(5), and 8.9(3) $\mu_B$, respectively, and the corresponding tilting angle from the $c$ axis is 26(1)$^\circ$, 28(1)$^\circ$, and 35(1)$^\circ$.}
\label{fig:magstr}
\end{figure*}

The individual spin-trimer chirality, as well as the entire chirality averaged over all sites, can be further tuned with a higher magnetic field of $\mu_0H>$ 1.2 T, where a new phase (III) is realized. In this phase, most of the magnetic reflections become weaker than those in phase II, and all the $(n/3\ n/3\ 0)$ reflections are extinct. These observations allow us to assign the magnetic structure shown in Fig.~\ref{fig:magstr}c, where the $c$ component of all the Dy$^{3+}$ moments aligns in the field direction~\cite{supp}. As shown in Fig.~\ref{fig:magstr}e, the in-plane components of the Dy$^{3+}$ moments change from all-in-all-out to tangential alignment, leading to two twisted trimer configurations with negative chirality of $-\chi_0'/3$ with $\chi_0'\approx 0.23$. Meanwhile, the twisted trimers also induce toroidal-like correlations over the second-neighbour triangles, where the averaged global chirality per spin-trimer unit amounts to $\chi_0'$, with a positive sign and an absolute magnitude that is three times as large as that over the nearest-neighbour spin trimers.

The global chirality in phases II and III motivated us to search for the chirality-induced GHE. Figure~\ref{fig:hall}  summarizes the results of our transport experiments. The anomalous Hall conductivity $\sigma_{xy}^{\mathrm{A}}$ was obtained by subtracting the normal Hall conductivity $\sigma_{xy}^{\mathrm{N}}$ from the total Hall conductivity $\sigma_{xy}^{\mathrm{tot}}$ with $\sigma_{xy}^{\mathrm{A}}=  \sigma_{xy}^{\mathrm{tot}}-\sigma_{xy}^{\mathrm{N}}$ \cite{supp}. The total Hall conductivity is related to the longitudinal resistivity $\rho_{xx}$ and the Hall resistivity $\rho_{yx}$ via $\sigma_{xy}^{\mathrm{tot}}= \rho_{yx}/(\rho_{xx}^2+ \rho_{yx}^2)$. In our definition, the anomalous Hall conductivity $\sigma_{xy}^{\mathrm{A}}$ may involve contributions not only from the conventional SOC-induced Hall term but also from the chirality-induced geometrical Hall term of the present focus.

As is shown in Figs.~\ref{fig:hall}a and b, the anomalous Hall conductivity $\sigma_{xy}^{\mathrm{A}}$ becomes non-zero in phases II and III as is expected for the chirality-induced GHE, and the evolution of $\sigma_{xy}^{\mathrm{A}}(H)$ loosely follows the magnetization curve $M(H)$ shown in the same panels. The successive step-like increases in $\sigma_{xy}^{\mathrm{A}}(H)$ across the phase transitions sharply contrast with the highly non-monotonous evolution of the longitudinal resistivity~\cite{supp}, indicating the independence of $\sigma_{xy}^{\mathrm{A}}$ on the relaxation time $\tau$ of the conducting electrons and thus excluding the extrinsic skew-scattering mechanism~\cite{nagaosa_anomalous_2010, onoda_quantum_2008, manyala_large_2004, miyasato_crossover_2007} that obeys $\sigma_{xy}^{\mathrm{skew}} \propto \tau$. Note also that the observed AHE reaches a Hall angle ($\sigma_{xy}^A/\sigma_{xx}$) as large as $1.5 \times 10^{-2}$ at 3 T in phase III.

\begin{figure*}[t!]
\includegraphics[width=0.64\textwidth]{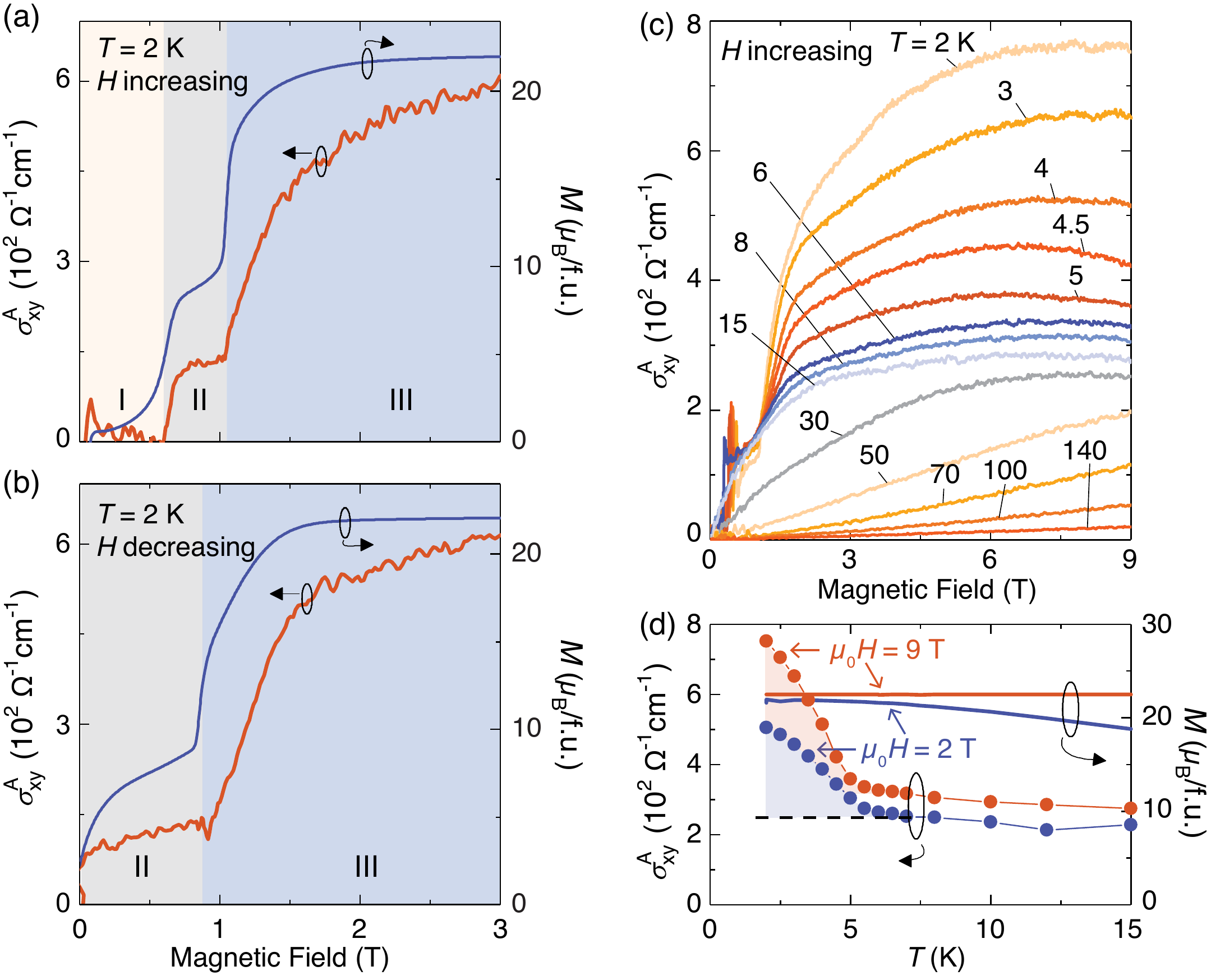}
\caption{(color online). (a,b) Anomalous Hall conductivity $\sigma_{xy}^{A}$ measured at 2 K as a function of increasing (a) and decreasing (b) magnetic fields. The corresponding field-dependence of magnetization $M(H)$ is shown for comparison. Different phases are indicated by color shading, and a large hysteresis for increasing and decreasing fields is observed at the phase boundaries. Note $\sigma_{xy}^{A}=0$ in phase I, where scalar spin chirality is absent, and a sharp increase of $\sigma_{xy}^{A}$ at the boundary between phases II and III. (c) Full view of $\sigma_{xy}^{A}$ data, including high fields and temperatures. (d) Temperature dependence of $\sigma_{xy}^{A}$ (filled circles) and magnetization (solid lines) measured in magnetic fields of 2 T (blue) and 9 T (red). A sharp increase in $\sigma_{xy}^{A} (T)$ when entering phase III below $\sim5$ K is in strong contrast to the behavior of $M(T)$ in the same field, evidencing the onset of non-zero scalar spin chirality at $T\sim 5$ K. Effect of demagnetization field has been corrected~\cite{supp}.
}
\label{fig:hall}
\end{figure*}

Two other possible scenarios for the observed AHE, including the intrinsic Karplus-Luttinger (momentum-space Berry phase) and the extrinsic side-jump mechanisms, can be excluded as the major origin on the basis of the temperature dependence of $\sigma_{xy}^{\mathrm{A}}$ shown in Figs.~\ref{fig:hall}c and d. In both mechanisms, $\sigma_{xy}^{\mathrm{A}}$ is independent of the relaxation time $\tau$~\cite{nagaosa_anomalous_2010, onoda_quantum_2008}. However, the magnitude of $\sigma_{xy}^{\mathrm{A}}$ induced through the Karplus-Luttinger or the side-jump mechanism is not related to chirality of the magnetic moments and only depends on the total magnetization $M$ along the field direction~\cite{nagaosa_anomalous_2010}. As is shown in Fig.~\ref{fig:hall}d, a sharp increase in $\sigma_{xy}^{\mathrm{A}}(T)$ is observed when the system is cooled through the phase transition at $\sim 5$~K in a magnetic field, while the magnetization $M(T)$ almost stays constant on both sides of the transition. The contrasting behavior of $\sigma_{xy}^{\mathrm{A}}(T)$ and $M(T)$ reveals that the observed AHE is not dominated by the Karplus-Luttinger or the side-jump mechanism and is consistent with its chirality origin.

The relative change of $\sigma_{xy}^A$ between phases II and III can be semi-quantitatively understood through the chirality-induced GHE. Assuming the geometrical weight over the Dy$^{3+}$ trimers and the second-neighbour triangles to be dominant and comparable~\cite{tatara_chirality_2002, onoda_anomalous_2004}, the total chirality $\chi^{\mathrm{tot}}$ of the Dy sublattice can be estimated to be $\sim 2\chi_0 /3$ and $2\chi_0' /3$ in phases II and III, respectively, with $\chi_0'/\chi_0\approx 1.4$. Meanwhile, due to the perturbative nature of the couplings between the conducting electrons and the Dy$^{3+}$ moments, the spin polarization $p$ of the conducting electrons should be proportional to the net magnetization, with $M^{\mu_0H = 2 \mathrm{T}}/M^{\mu_0H = 0.7 \mathrm{T}} \approx 2.7$. Given that $\sigma_{xy}^{\mathrm{A}} \propto p\chi_{\mathrm{tot}}$, the ratio of the chirality-induced anomalous Hall conductivity in phases III and II is expected to be $\sim 3.8$, which is close to our observation in Fig.~\ref{fig:hall}a.

Compared to the skyrmion lattice characterized by a topological Hall effect related to the winding number of the spin texture~\cite{lee_unusual_2009, ritz_formation_2013, nagaosa_topological_2013,kurumaji_skyrmion_2019, hirschberger_skyrmion_2018}, the magnetic structures in phases II and III of Dy$_3$Ru$_4$Al$_{12}$ have a much shorter periodicity, where a magnetic unit cell consists only of $3\times3$ unit cells. According to theoretical calculations~\cite{onoda_anomalous_2004}, when the couplings between the conducting electrons and local moments are weak as in Dy$_3$Ru$_4$Al$_{12}$, the two Berry-phase scenarios in reciprocal and real spaces can be equivalent. Therefore, Dy$_3$Ru$_4$Al$_{12}$ might be viewed as an intermediate system that bridges the magnetic skyrmion lattice with a long periodicity of tens of nanometers~\cite{nagaosa_topological_2013} and the prototypical kagom\'e model with a minimal periodicity of one single unit cell~\cite{ohgushi_spin_2000}, and might help clarify the equivalence between the two Berry-phase scenarios for the GHE in reciprocal and real spaces. 

Due to the strong SOC on the Ru ions, we do not completely rule out the role of SOC in our observed GHE. The SOC may join the scalar spin chirality in enhancing the transfer of the Berry curvature. Recent investigations on Mn$_3$Sn and Mn$_3$Ge reveal that the SOC is able to induce an AHE even in antiferromagnets with coplanar magnetic structures~\cite{chen_anomalous_2014,nakatsuji_large_2015, nayak_large_2016,liu_anomalous_2017}. It will be interesting to clarify the role of the SOC and explore its possible interplay with the scalar spin chirality in Dy$_3$Ru$_4$Al$_{12}$.

In summary, our neutron diffraction experiments reveal the existence of spin trimers in the breathing kagom\'e lattice compound Dy$_3$Ru$_4$Al$_{12}$, where both the stacking order and the local scalar spin chirality of the trimers can be tuned by a magnetic field. In phases with non-zero entire scalar spin chirality, a large GHE is observed in our magneto-transport experiment. Our works provide an unambiguous illustration for the chirality induced GHE, and reveal that systems composed of tunable spin trimers can exhibit large emergent electromagnetic response due to couplings between the conduction electrons and the localized magnetic moments.

\begin{acknowledgments}
We acknowledge helpful discussions with H. Ishizuka, N. Nagaosa, N. Gauthier, B. Normand, Owen Benton, and C.L. Zhang.  Our neutron diffraction experiments were performed at the Swiss Spallation Neutron Source SINQ, Paul Scherrer Insitut PSI, Villigen, Switzerland. Our transport experiment in high magnetic fields up to 24 T were measured at the High Field Laboratory for Superconducting Materials at Tohoku University. This work was supported in part by JST CREST Grant Number JPMJCR1874 (Japan). M. H. was supported as a JSPS International Research Fellow (18F18804).
\end{acknowledgments}



\renewcommand{\thefigure}{S\arabic{figure}}
\renewcommand{\thetable}{S\arabic{table}}

\renewcommand{\theequation}{\arabic{equation}}

\makeatletter
\renewcommand*{\citenumfont}[1]{S#1}
\renewcommand*{\bibnumfmt}[1]{[S#1]}
\makeatother

\setcounter{figure}{0} 
\setcounter{table}{0}
\setcounter{equation}{0} 

\onecolumngrid
\newpage
\begin{center} {\bf \large Ordering phenomena of spin trimers accompanied by large geometrical Hall effect \\
 Supplementary Information} \end{center}
\vspace{0.5cm}
\onecolumngrid


 \section{I. Crystal growth and refinement of the x-ray diffraction data}
 
 Dy$_3$Ru$_4$Al$_{12}$ single crystals were grown using the Czochralsky technique with 1~\% excess Al in raw ingots to compensate evaporation during the melt-growth process \cite{gorbunov_electronic_2014s}. Phase-purity and absence of grain boundaries in the resulting crystals were confirmed using powder x-ray diffraction (XRD), energy-dispersive x-ray spectroscopy (EDX), scanning electron microscopy (SEM), as well as optical microscopy. 
 
\begin{figure}[h!]
\includegraphics[width=0.60\textwidth]{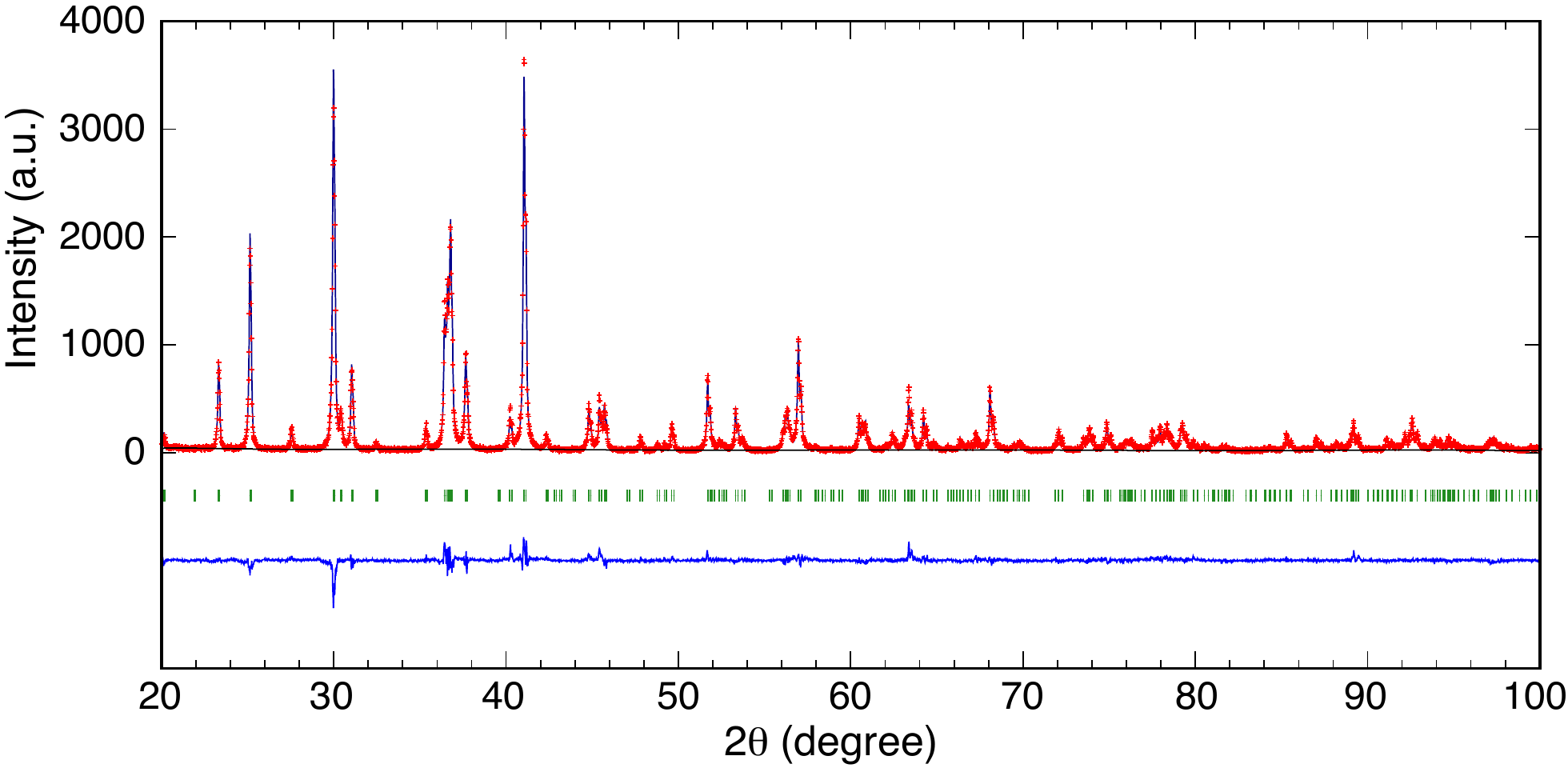}
\caption{Rietveld refinement results of the x-ray diffraction (XRD) data for Dy$_3$Ru$_4$Al$_{12}$ powder. Data points are shown by red crosses. The calculated pattern is shown by the blue solid line. Vertical bars show the positions of the Bragg peaks. The blue line at the bottom shows the difference between data and calculated intensities.
}
\label{fig:xrd}
\end{figure}

XRD measurements on pulverized Dy$_3$Ru$_4$Al$_{12}$ crystals were performed with a commercial in-house x-ray diffractometer (Rigaku RINT TTR-III, Cu $K_{\alpha}$ radiation) at room temperature. Refinements were performed with the space group  $P6_3/mmc$ using the software RIETAN~\cite{izumi_three_2007}. The refinement results are shown in Fig.~\ref{fig:xrd}.  The refined lattice constants are $a=8.774(2)$ \AA, and $c=9.530(1)$~\AA. The refined Dy position at the $6h$ site $(x\ 2x\ 1/4)$ is $x =0.1934(1)$. It was necessary to take into account a preferred orientation of $(001)$ plane with a preference factor of $1.10$~\cite{izumi_three_2007}. The goodness-of-fit $R$ factors are $R_{wp}=13.0\,\%$, $R_p = 9.8\,\%$. 

 \section{II. Magnetization measurements and phase diagram}
\begin{figure}[h!]
\includegraphics[width=0.79\textwidth]{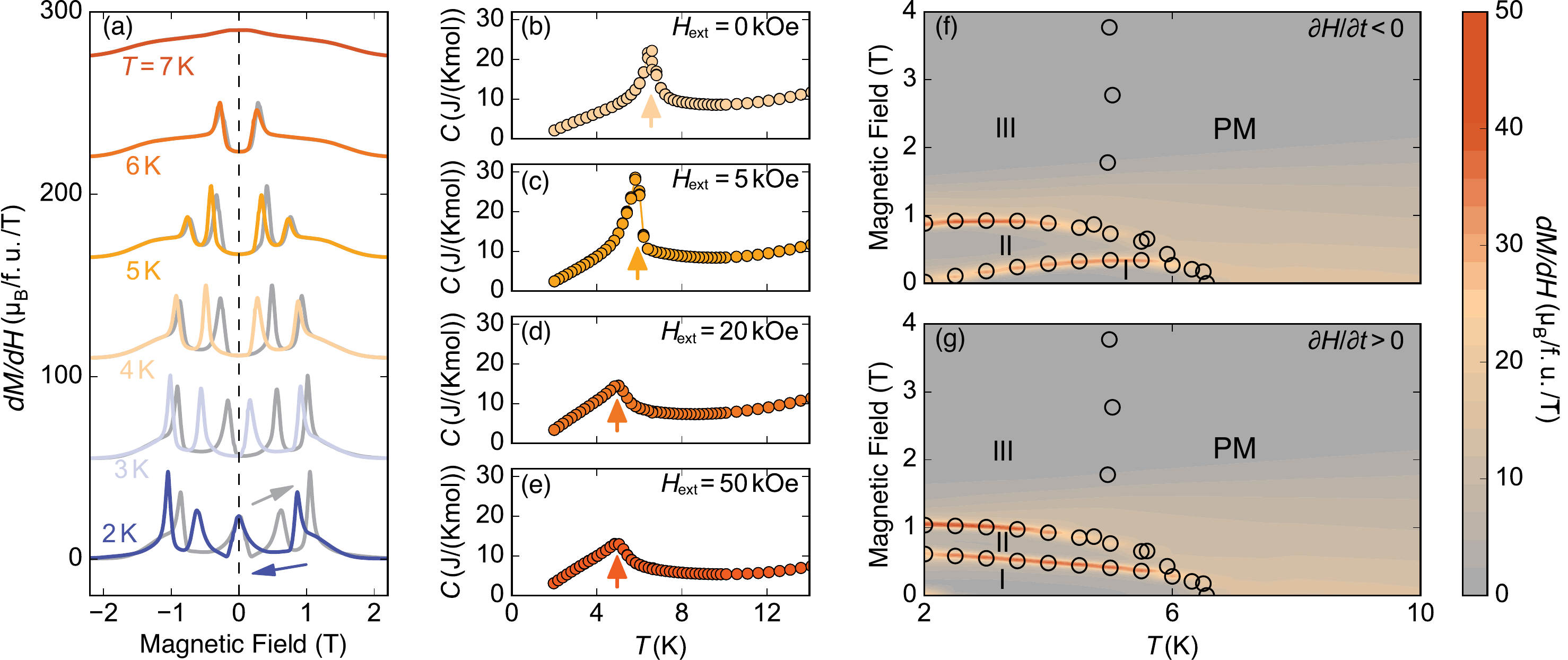}
\caption{(a) Magnetic susceptibility, (b-e) heat capacity (measured in cooling), and (f,g) phase diagram in a magnetic field along the $c$ axis. We constructed  magnetic phase diagrams in panels (f) and (g) with phases I, II, and III as specified in the main text, using magnetic susceptibility and specific heat measurements. PM signifies the paramagnetic state. The data in (a) are offset by constant values for clarity. Data for decreasing (increasing) magnetic field is shown in color (grey). In (b-e), we identify a sharp first-order transition at lower fields ($H=0\ \mathrm{and}\ 5\ $kOe). Although the transition from phase III to the field-aligned PM state is invisible in the magnetization data, the specific heat measurements clearly delineate the two regimes as indicated by orange arrows. In panels (f) and (g), $\partial H/\partial t <0$ ($\partial H/\partial t > 0$) signifies decreasing (increasing) magnetic field. In panels (a,f,g), effect of the demagnetization field has been corrected.
}
\label{fig:pd}
\end{figure}

Magnetization and specific heat measurements were employed to characterize the phase diagram of Dy$_3$Ru$_4$Al$_{12}$ in a magnetic field along the $c$ axis. These data agree with the transport and neutron scattering experiments discussed in the main text, if the effect of the demagnetization field is corrected by $H = H_\text{ext} - NM$ with internal field $H$, external field $H_\text{ext}$, magnetization $M$, and the averaged demagnetization factor $N$. The demagnetization factor was estimated for both cuboids and platelets by elliptical approximation of the sample shape~\cite{osborn_demagnetization_1945}. The derivative $d M/d H$ in Fig.~\ref{fig:pd}a was obtained from the magnetization data $M(H)$ measured with a vibrating sample magnetometer (VSM) in a Quantum Design PPMS cryostat. We evidenced two field-induced magnetic transitions with a large hysteresis. At the lowest temperature, the transition between phases I and II (c.f. main text) occurs very close to zero magnetic field in field-decreasing measurements, i.e. phase II can be meta-stabilized down to very low field (Fig.~\ref{fig:pd}f). 

Although phase III and the field-aligned paramagnetic regime (PM, Figs.~\ref{fig:pd}f and g) are indistinguishable in magnetization measurements at high fields, a sharp transition between them was observed in the specific heat measurements (Fig.~\ref{fig:pd}d and e). The thermodynamic measurement is thus in agreement with the previous report on the $T$ and $H$ dependence of the ultrasound velocity in Dy$_3$Ru$_4$Al$_{12}$.~\cite{ishii_magnetic_2018s}.

 \section{III. Refinement of the neutron diffraction dataset}
 
Neutron diffraction experiments on a single crystal sample of Dy$_3$Ru$_4$Al$_{12}$ were performed on the thermal-neutron diffractometer ZEBRA at Swiss Spallation Neutron Source SINQ of Paul Scherrer Institut PSI (Villigen, Switzerland). A piece of Dy$_3$Ru$_4$Al$_{12}$ single crystal with dimensions of 3.88$\times$3.71$\times$1.40 mm$^3$ was aligned with the $(hk0)$ plane horizontal. A cryomagnet with vertical magnetic field and base temperature of 2.2 K was employed. Incoming neutron wavelength of 1.18 \AA\ (Ge(311) monochromator) was selected for the measurements. At zero field, 266 magnetic reflections, of which 190 are mutually independent, were collected. At an external magnetic field of 0.7 (2.0) T, 230 (64) independent magnetic reflections were collected. Absorption corrections of the intensities were performed using JANA~\cite{petricek_crystallographic_2014}. Refinements of the datasets were carried out using FULLPROF \cite{rodriguez_recent_1993}. 
 
\subsection{Phases I and II}
\begin{figure}[h!]
\includegraphics[width=0.62\textwidth]{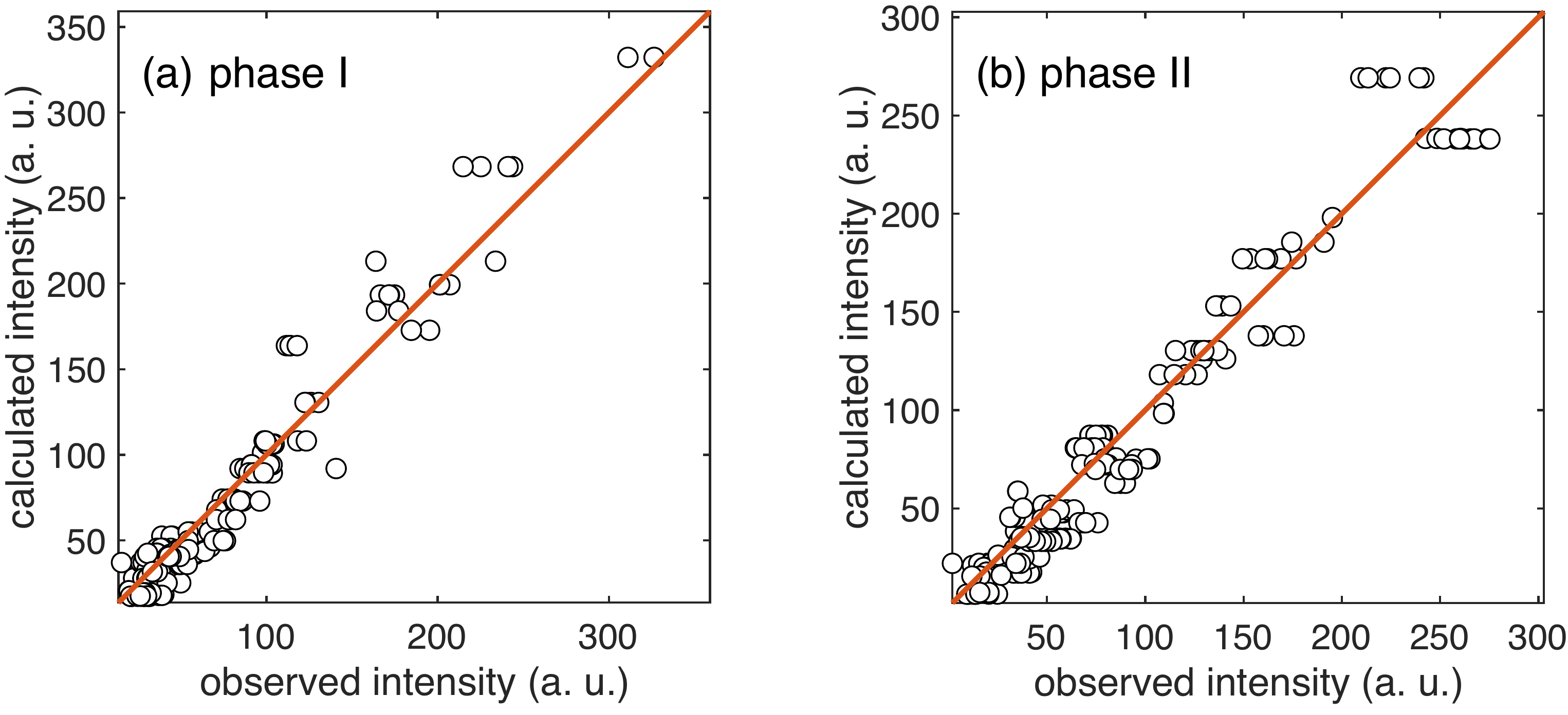}
\caption{Comparison of the observed and calculated intensities for the refined magnetic structures in phases I (a) and II (b).
}
\label{fig:refine_phase12}
\end{figure}

Refinements of the neutron diffraction datasets were performed with FullProf~\cite{rodriguez_recent_1993}. Following Ref.~\cite{gorbunov_electronic_2014}, a satisfactory fit for the dataset collected in zero field was obtained with the magnetic space group $C_c2/c$ (BNS No. 15.90). The refined magnetic structure is similar to that reported in Ref.~\cite{gorbunov_electronic_2014}, and the $R$-factors are $R_f=10.3\,\%$ and $R_{f2}=16.1\,\%$. However, under further constraints of equal moment size and all-in-all-out alignment for regular spin trimers, we were able to obtain a similarly good fit, with comparable $R$-factors of $R_f=10.1\,\%$ and $R_{f2}=16.1\,\%$. The refined magnetic structure is shown in Fig.~2a of the main text and the comparison of the observed and calculated intensities is shown in Fig.~\ref{fig:refine_phase12}a.

For the refinement of the neutron diffraction dataset collected in phase II, we started with the spin configuration observed in phase I for the primary unit cell and introduced a sinusoidal modulation with $\bm{q}_2=$ (1/3 1/3 0). In this way, a relatively good fit was obtained with $R$-factors of $R_f=10.6\,\%$ and $R_{f2}=16.8\,\%$. However, this solution involves unequal magnetic moment sizes even after adding a uniform spin component along the $c$ axis that is induced by the applied field. To overcome this problem, we introduced a similar magnetic structure that shows an up-up-down sequence for the spin trimers along $a$ and $b$ axes, which can be viewed as a modified version of the sinusoidally modulated structure. This solution has equal moment size and is able to explain the magnetization plateau observed in phase II. Refinement against the neutron diffraction dataset results in a satisfactory fit with $R$-factors of $R_f=9.8\,\%$ and $R_{f2}=15.6\,\%$, and the comparison for the observed and calculated intensities is shown in Fig.~\ref{fig:refine_phase12}b.

\subsection{Phase III}
As is mentioned in the main text, intensities of the magnetic reflections belonging to $\bm{q}_2 =$ (1/3 1/3 0) are weaker in phase III compared to those in phase II, indicating that the $c$ components of the magnetic moments are aligned parallel to the applied field and are thus forming the $\bm{q}=0$ propagation vector. Therefore, only the in-plane components contribute to the $\bm{q}_2=$ (1/3 1/3 0) reflections, leading to weaker intensities. This in-plane character has also been unambiguously proved in our recent resonant x-ray diffraction experiment~\cite{dra_rxd}.

\begin{figure}[b!]
\includegraphics[width=0.75\textwidth]{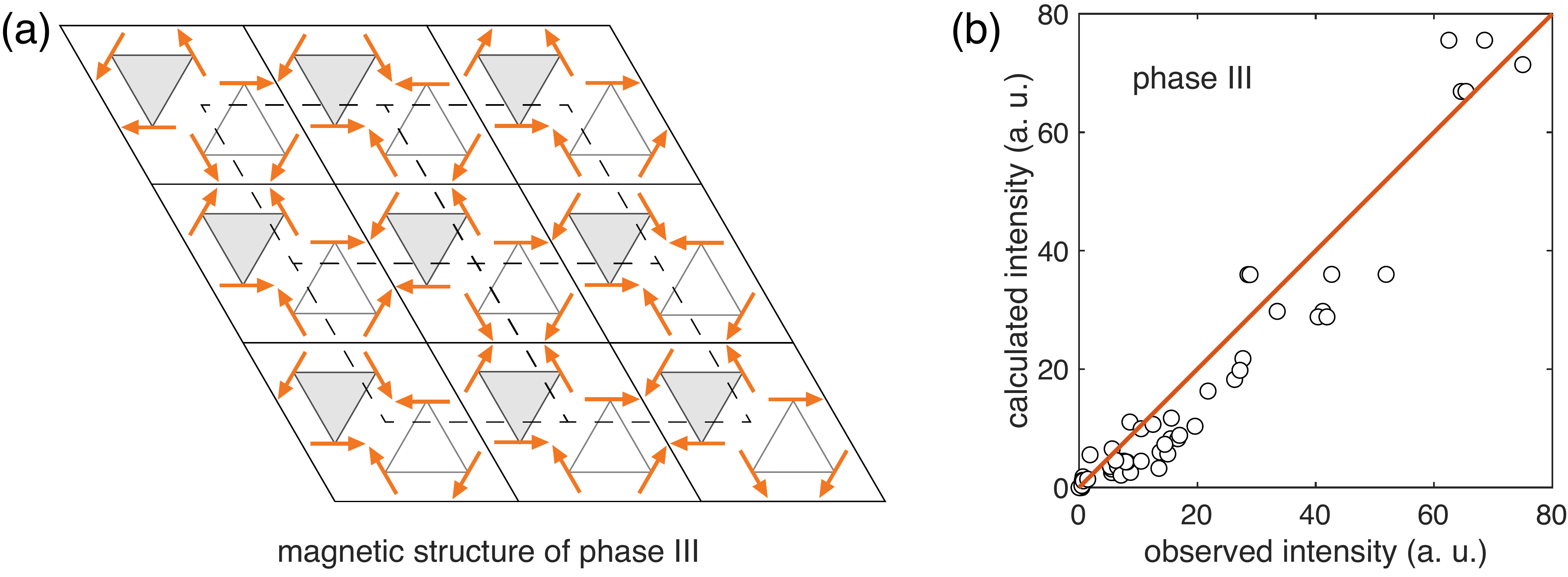}
\caption{(a) In-plane components of the refined magnetic structure in phase III. Dashed lines indicate the unit cell shifted by (1/2 1/2 0) to emphasize the hexagon formed by Dy$^{3+}$ ions in two neighbouring layers. (b) Comparison of the observed and calculated intensities for the refined magnetic structure in phase III.
}
\label{fig:refine_phase3}
\end{figure}

Magnetic structure of the $\bm{q}_2 =$ (1/3 1/3 0) component in phase III was determined through symmetry analysis. It is known that most commensurate spin configurations possess one of the possible maximal magnetic symmetries compatible with the propagation vector in the parent space group, forming the k-maximal subgroups~\cite{perez_symmetry_2015}. Using the program \textit{k-Subgroupsmag} in the Bilbao Crystallographic Server~\cite{bilbao}, we obtained 9 k-maximal subgroups of the $P6_3/mmc$ parent space group with k = (1/3 1/3 0), among which $P6_3/m'c'm'$ (BNS No.~193.261), $P6_3'/m'cm'$ (193.259), $P6_3'/m'c'm$ (193.258), $P6_3/m'cm$ (193.255), $P\overline{6'}2m'$ (189.224), $P\overline{6'}2'm$ (189.223) allow spin components in the $ab$ plane. Refinement of the neutron diffraction dataset reveals that the best fit is achieved with the $P6_3'/m'c'm$ (193.258) magnetic space group with $R$-factors of $R_f=19.1\,\%$ and $R_{f2}=25.9\,\%$. The corresponding magnetic structure is shown in Fig.~2c of the main text, and is also reproduced in Fig.~\ref{fig:refine_phase3} together with the comparison of the observed and calculated intensities. The relatively large value of the $R$-factors compared to those in phases I and II might due to the relatively weaker intensities of the magnetic reflections in phase III.

\begin{figure}[h!]
\includegraphics[width=0.55\textwidth]{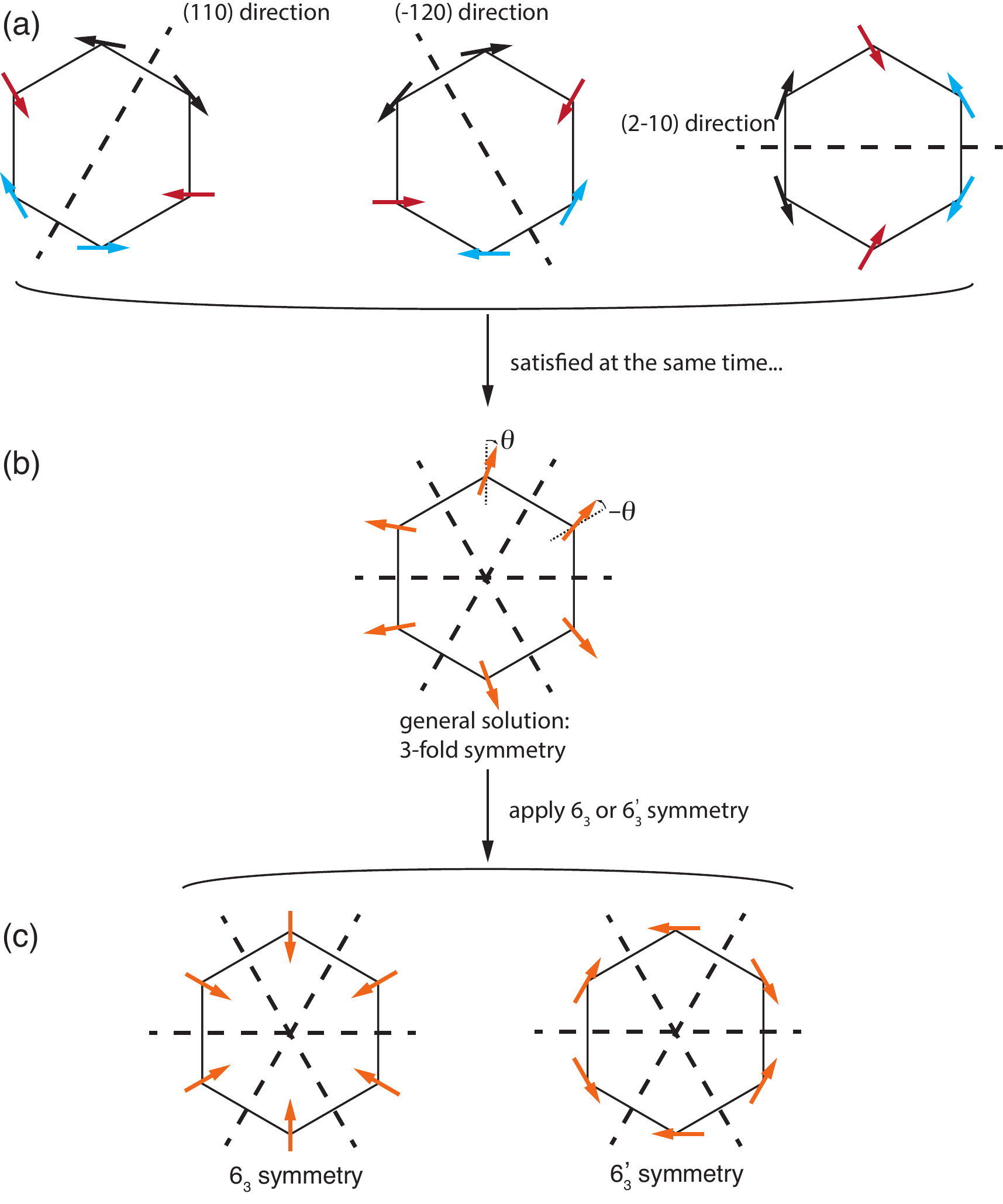}
\caption{Diagram for the solution of the magnetic structure in phase III using the extinction rule. (a) The observed $(n/3\ n/3\ 0)$ extinction rule constrains the spin alignment over the Dy$^{3+}$ hexagons to be symmetrical with respect to the $\langle110\rangle$ directions. (b) Constraints along all three $\langle110\rangle$ directions leave only one degree of freedom in the spin alignment, that is the rotation around the $c$ axis with a degree of $\pm\theta$. (c) Applying the $6_3$ or $6'_3$ symmetry leaves only two possible solutions for the spin alignment, corresponding to $\theta_1 = 0$/$\pi$ and $\theta_2 = \pm0.5 \pi$, respectively.
}
\label{fig:refine_solution}
\end{figure}

The magnetic structure in phase III was also confirmed with a different method that is based on the observed extinction rule. Among the $(n/3\ n'/3\ 0)$ magnetic reflections, zero intensity is observed for $n=n'$. Considering that neutron scattering only probes the magnetic structure factor component that is perpendicular to the momentum transfer, this extinction rule means that for any $(n/3\ n/3\ 0)$ momentum transfer, the magnetic structure factor is either zero or parallel to the $\langle110\rangle$ directions, where $\langle110\rangle$ represents the symmetrically equivalent directions along $\bm{a}$, $\bm{b}$, and $\bm{a}+\bm{b}$. Therefore, in the unit cell shifted by (1/2 1/2 0) as indicated by dashed lines in Fig.~\ref{fig:refine_phase3}a, the Dy$^{3+}$ moments over the hexagon should be symmetric with respect to the $\langle110\rangle$ directions as illustrated in Fig.~\ref{fig:refine_solution}a. Constraints along the three $\langle110\rangle$ directions leave only one degree of freedom in the spin alignment, that is the rotation of the spins by an angle of $\pm \theta$ shown in Fig.~\ref{fig:refine_solution}b, where the $\pm$ sign differentiates the spins in different layers. By further applying the $6_3$ or $6'_3$ symmetry, we are able to obtain the two special solutions with $\theta_1 = 0$/$\pi$ and $\theta_2 = \pm 0.5\pi$ shown in Fig.~\ref{fig:refine_solution}c. Based on these two candidate alignments within the unit cell, we could construct the $\bm{q}_2 =$ (1/3 1/3 0) magnetic structure on the whole lattice. The solution with $\theta_2 = \pm 0.5\pi$ is exactly the same as what we found through magnetic space group analysis, while the solution with $\theta_1 = 0$ or $\pi$ leads to a worse fit with $R_{F2} = 85.27\ \%$.

\section{IV. Stability of the spin trimers and phase transitions}
The rigid spin trimers in phases I and II revealed in our neutron diffraction experiment allow us to understand the meta-magnetic phase transitions in Dy$_3$Ru$_4$Al$_{12}$. Noticing the relatively strong ferromagnetic nearest-neighbour (NN) coupling $J_1$ that is evidenced by the positive Weiss temperature of $\sim55$ K, we can treat the spin trimers as effective Ising spins with only $c$ component, which simplifies the breathing kagom\'e lattice of the Dy$^{3+}$ spins as an effective triangular lattice of composite Ising spins. Here we neglect the couplings between the neighbouring layers in the $c$ direction, which is of antiferromagnetic character on the basis of the $q_{c}=1/2$ modulation along the $c$ axis in phase I but should be weak since $q_{c}$ becomes zero in phase II at a relatively weak magnetic field of  0.6 T.

\begin{figure}[t]
\includegraphics[width=0.45\textwidth]{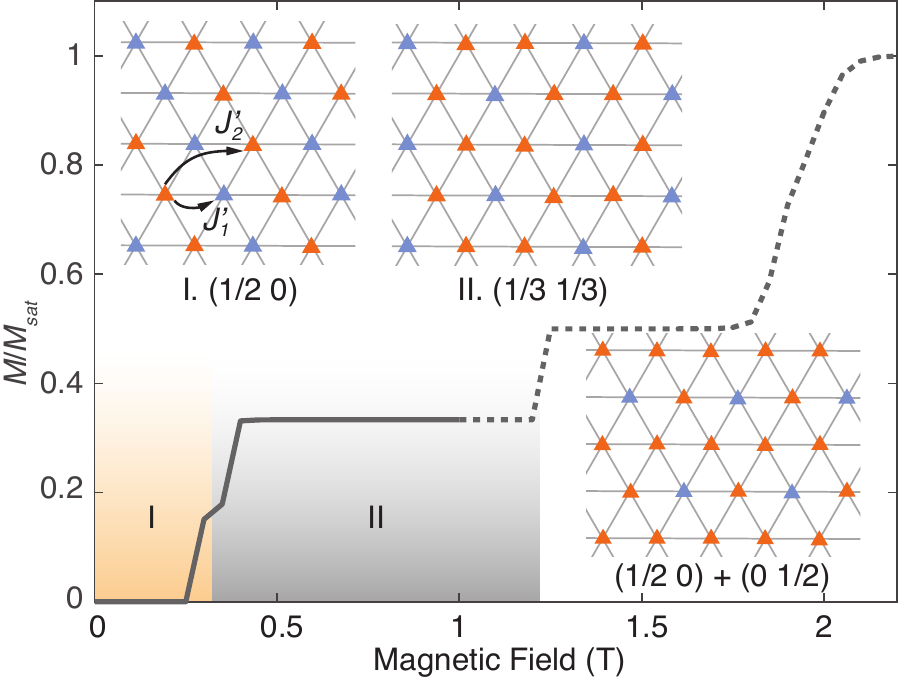}
\caption{Monte Carlo simulation with spin trimers treated as composite Ising spins. Field-dependence of magnetization for the triangular lattice with spin trimers treated as classical Ising spins. The simulation was performed on a 12$\times$12 superlattice with $J'_1=0.38$ K, $J'_2=J'_1/10 = 0.038$~K, and $T= 2$ K. The magnitude of the magnetic moment is $22$ $\mu_B$, which equals the saturated moment of the Dy$^{3+}$ trimer in a high magnetic field along the $c$ direction. The dashed part of the magnetization curve indicates the high-field region where the rigid spin trimers break down. The narrow plateau on the boundary of phases I and II is due to the finite size effect. The insets show the alignment of the composite Ising spins in phase I with $\bm{q}'=$ (1/2 0) at the top left, phase II with $\bm{q}' =$ (1/3 1/3) at the top right, and phase III with double $\bm{q}'$ of (1/2 0) + (0 1/2) at the bottom. In each alignment, red (blue) triangles represent spin trimers with $c$ component aligned parallel (anti-parallel) to the field.
}
\label{fig:monte_carlo}
\end{figure}

The phase diagram for the triangular lattice with classical Ising spins was investigated previously~\cite{metcalf_ground_1974}. It is known that in a large parameter space of antiferromagnetic NN couplings $J_1'$ and second-neighbour couplings $J_2'$, the zero-field ground state has a propagation vector of $\bm{q}' =$ (1/2 0) as we observed in phase I of Dy$_3$Ru$_4$Al$_{12}$. In a magnetic field perpendicular to the triangular lattice plane, a series of magnetic transitions appears, and the evolution of $\bm{q}/(H)$ depends on the ratio of $J_2'/J_1'$. Specifically, as is shown in Fig.~\ref{fig:monte_carlo}, when $J_2'/J_1' < 0.2$, an intermediate phase with $\bm{q}'=$(1/3 1/3) emerges as we observed in phase II of Dy$_3$Ru$_4$Al$_{12}$, after which the system crosses a double-$\bm{q}$ phase with $\bm{q}'=$(1/2 0) and (0 1/2), before finally settling in the fully polarized phase with $\bm{q}'=0$. The absence of the latter two phases in Dy$_3$Ru$_4$Al$_{12}$ indicates the breakdown of rigid spin trimers in a field above $\sim 1.1$ T, which might be ascribed to the crossing of the Dy$^{3+}$ crystal field levels that have been revealed in the ultrasound experiment~\cite{ishii_magnetic_2018}.

\begin{figure}[b!]
\includegraphics[width=0.69\textwidth]{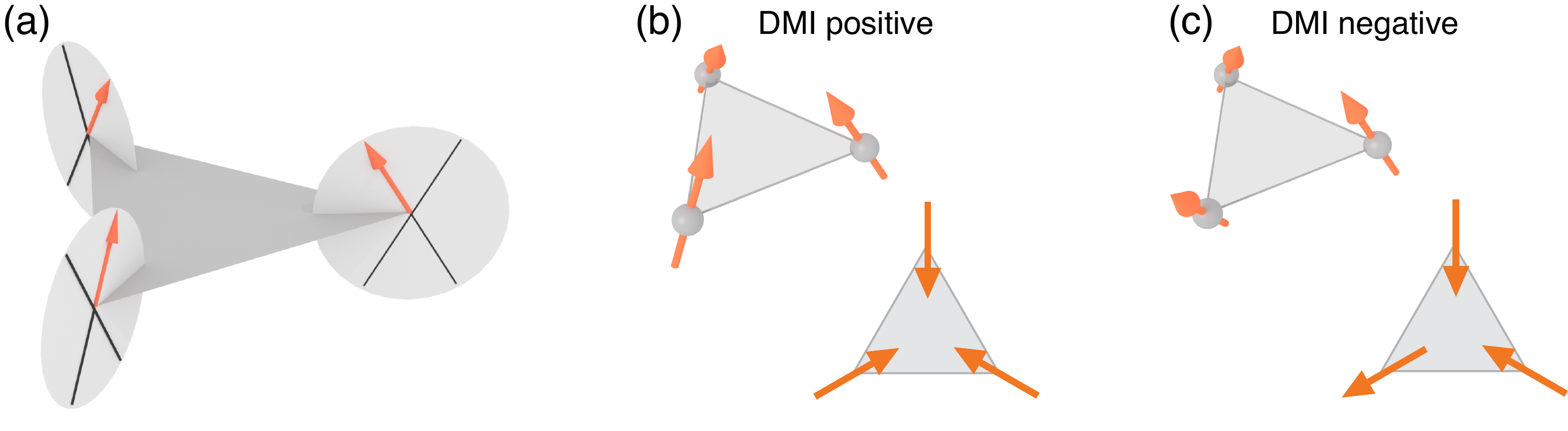}
\caption{Anisotropy of the $R^{3+}$ magnetic moment and the spin trimer configurations. (a) Spin trimers in $R_3$Ru$_4$Al$_{12}$ with the mirror plane shown as gray planes. Solid lines in the circles indicate the two symmetric easy axes. The $R^{3+}$ moments aligning along the easy axes might realize the 4-state clock model. Panels (b) and (c) show the spin trimer configurations under positive and negative DMI interactions, respectively. Directions of the DM vectors are defined in Fig.1b of the main text. 
}
\label{fig:potts}
\end{figure}

\section{V. Single ion anisotropy of the rare earth ions in R$_3$Ru$_4$Al$_{12}$}
Besides Dy$_3$Ru$_4$Al$_{12}$, the magntic properties the $R_3$Ru$_4$Al$_{12}$ family with $R = $ La $\sim$ Nd, Gd $\sim$ Er, and Yb have been studied~\cite{ge_physical_2014, henriques_complex_2018, gorbunov_magnetic_2016, hirschberger_skyrmion_2018s, nakamura_spin_2018, rayaprol_neutron_2019, gorbunov_crystal_2018,upadhyay_magnetic_2017, rayaprol_neutron_2019, nakamura_low_2015}, and a systematic evolution of the single ion anisotropy of the $R$ ions can be extracted. For $R=$ Pr~\cite{henriques_complex_2018}, Nd~\cite{gorbunov_magnetic_2016}, Tb~\cite{rayaprol_neutron_2019}, Dy, and Ho~\cite{gorbunov_crystal_2018}, the magnetization along the $c$ axis is much larger than that within the $ab$ plane, and in their magnetic structures, if available from neutron diffraction, the $R^{3+}$ moments have a large component along the $c$ axis. All these $R^{3+}$ ions have negative Stevens factor $\alpha$~\cite{jensen_rare_1991}. By contrast, for $R$ = Yb~\cite{nakamura_low_2015} with positive Stevens factor $\alpha$, the magnetization along the $c$ axis becomes much smaller than that within the $ab$ plane. This comparison reveals that the single ion anisotropy of the $R^{3+}$ ions in $R_3$Ru$_4$Al$_{12}$ is dominated by the Stevens factor $\alpha$, similar to the situation in many rare earth oxides~\cite{huang_quantum_2014}.

However, it should be noted that an overall easy axis along the $c$ axis in magnetization does not necessarily mean that the local easy axis of the $R^{3+}$ magnetic moment should also be exactly along the $c$ axis. In $R_3$Ru$_4$Al$_{12}$, the $R^{3+}$ ions occupy the 6$h$ site with 2$mm$ site symmetry. As is shown in Fig.~\ref{fig:potts}a, one mirror plane is the $ab$ plane, and the other mirror plane is perpendicular to the $ab$ plane and contains the $a^*$ direction in reciprocal space. Therefore, in the general case, the $R^{3+}$ ions can exhibit complicated anisotropy that involves two symmetric easy axes and thus realize the 4-state clock model. The titled easy axes might explain why the magnetization seems to be saturated at $\sim 2$ T in phase III but stays below the expected value even in a high field up to 60 T~\cite{gorbunov_electronic_2014}. 

Based on this special anisotropy, positive Dzyaloshinskii-Moriya Interactions (DMI) over the nearest-neighbour bonds (see Fig.~1 in the main text for the definition of the positive DMI vector direction) will lead to the all-in-all-out spin trimer configurations observed in Dy$_3$Ru$_4$Al$_{12}$, while negative DMI will result in either 2-in-1-out or 1-in-2-out configurations similar to that expected for the kagome ices~\cite{fennell_pinch_2007}. 

It should be noted that besides the single-ion anisotropy, anistropic couplings like dipolar interactions or anistropic exchange interactions might also contribute to the formation of spin trimers in Dy$_3$Ru$_4$Al$_{12}$. One possible scenario is that the anisotropy of the Dy$^{3+}$ magnetic moment is tilted but stays rotationally invariant around the $c$ axis, leading to a diabolo shape for the spin anisotropy. In this way, positive DM interactions will also favor the formation of all-in-all-out spin trimer configurations as observed in phases I and II. Further study will be needed to clarify the single ion anisotropy of the Dy$^{3+}$ ions and clarify the origin of the spin trimers.

\begin{figure}[h!]
\includegraphics[width=0.55\textwidth]{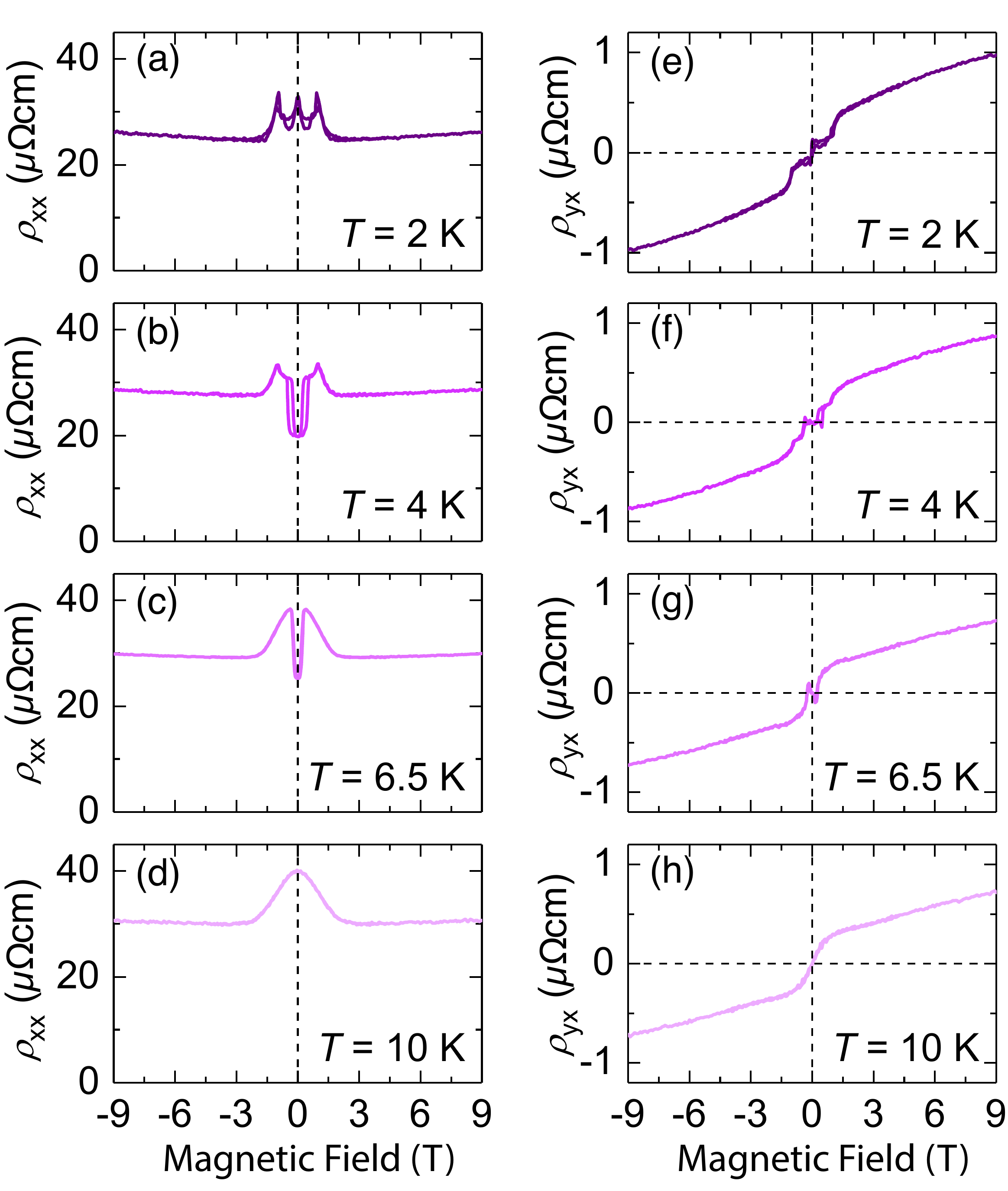}
\caption{(a-d) Longitudinal resistivity $\rho_{xx}$ and (e-h) Hall resistivity $\rho_{yx}$ of Dy$_3$Ru$_4$Al$_{12}$ as a function of magnetic field at selected temperatures. Low-field magnetic phase transitions are accompanied by a large hysteresis between field-increasing and decreasing ramps.
}
\label{fig:trans}
\end{figure}

\section{VI. Raw data of longitudinal and Hall resistivities}

In-plane longitudinal and Hall conductivities $\sigma_{xx}$ and $\sigma_{xy}$, as shown in the main text, were constructed from longitudinal and Hall resistivities $\rho_{xx}$ and $\rho_{yx}$ (Fig.~\ref{fig:trans}) using the well-known relations for the tensor elements:
\begin{align}
\sigma_{xx} &= \rho_{xx}/\left(\rho_{xx}^2+\rho_{yx}^2\right)\\
\sigma_{xy} &= \rho_{yx}/\left(\rho_{xx}^2+\rho_{yx}^2\right)
\end{align}
The sample geometry was carefully measured using an optical microscope (error bar $\pm5\,\%$) so as to minimize uncertainties of the analysis. Above $T_N$, the system shows strong negative magnetoresistance, indicating significant correlation between the conducting electrons and the short-range ordered magnetic moments. The more complex behavior of $\rho_{xx}$ and $\rho_{yx}$ below $T_N$ is related to the magnetic ordering, and consequent changes in carrier scattering time $\tau$. In such cases of field dependent $\tau$, it is preferable to focus on $\sigma_{xy}$ rather than of $\rho_{yx}$ for a proper analysis of the anomalous Hall or geometrical Hall effect.

However, the Hall resistivity $\rho_{yx}$ comes with the considerable advantage that the normal Hall effect due to the Lorentz force acting on electrons and holes has weak temperature dependence in the metallic state in general, as is also the case for the $R_3$Ru$_4$Al$_{12}$ family~\cite{hirschberger_skyrmion_2018}. Therefore, we first subtracted the field-linear normal Hall resistivity from $\rho_{yx}$ viz.
\begin{equation}
\rho_{yx}^A = \rho_{yx}-\rho_{yx}^N = \rho_{yx}-\rho_{yx}^N(300 \,\text{K})\text{,}
\end{equation}
and subsequently calculated the anomalous/geometrical Hall conductivity following $\sigma_{xy}^A = \rho_{yx}^A /\left(\rho_{xx}^2+\rho_{yx}^2\right)$. Note that the contribution of $\rho_{yx}$ in the denominator amounts to less than $10^{-3}$ in relative terms. 

Our simple analysis, which does not make any major ad-hoc assumptions about the behavior of the normal Hall effect, satisfies a simple `sanity check': In phase I with antiferromagnetic stacking of alternating magnetic layers and vanishing global scalar spin chirality, $\sigma_{xy}^A=0$ exactly within our analysis as is shown in Fig.~3 of the main text.

\section{VII. High-field transport experiments}

\begin{figure}[b!]
  \includegraphics[width=0.95\linewidth]{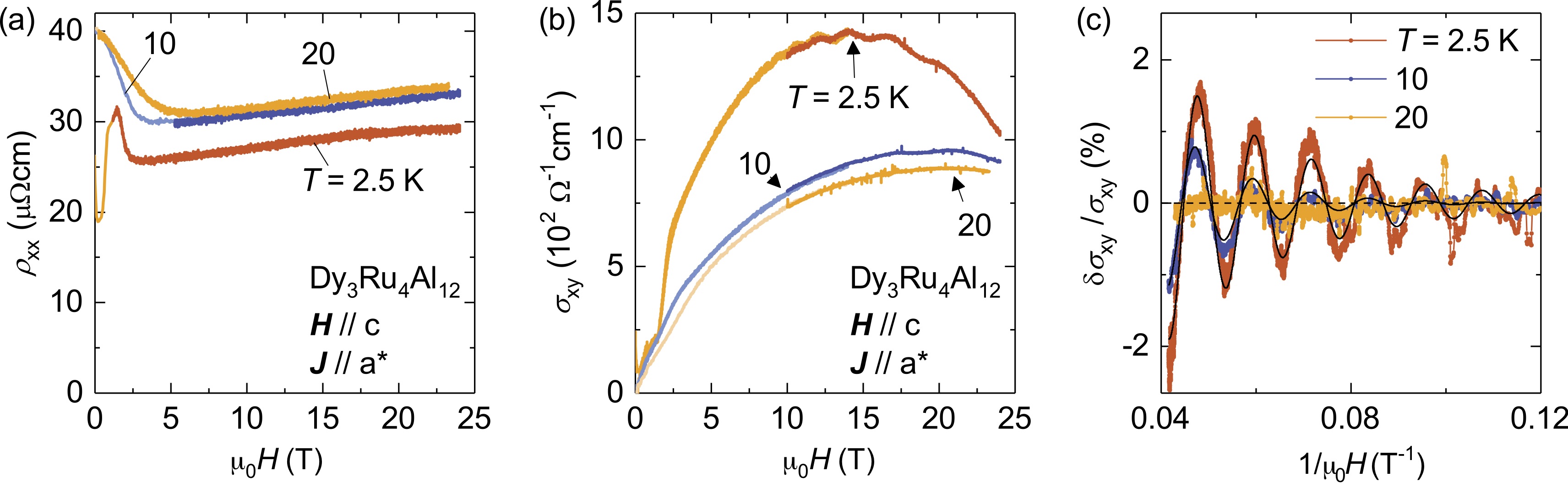}
  \caption{\textbf{Magneto-transport properties at higher fields.} (a) High-field magnetoresistance $\rho_{xx}(H)$ at three selected temperatures. In lighter colors, we show data obtained on the same crystal in an in-house measurement below $14\,$T. Sweep direction was $\partial H/\partial t<0$, $>0$, and $<0$ for $T=2.5\,$K, $10\,$K, and $20\,$K, respectively. (b) Hall conductivity $\sigma_{xy}$ measured under the same conditions. (c) We extracted the oscillatory part of the Hall conductivity and fitted to the Lifshitz-Kosevich expression (black lines). Fit parameters are provided in the text. Demagnetization correction was not applied to this data. High field data for $\rho_{xx}$ and $\rho_{yx}$ were corrected to match the low-field result, see text for details.}
\label{fig:highfield}
\end{figure}

At higher magnetic fields, it is expected that a closing of the spin-trimer `umbrella' structure leads to a suppression of the geometrical Hall signal. We explored the regime up to $\mu_0 H=24\,$T at the High Field Laboratory for Superconducting Materials (IMR, Tohoku University, Sendai, Japan) with a standard transport lock-in technique. The 25T cryogen-free superconducting magnet (25T-CSM) was used~\cite{awaji_first_2017}. The electric current was applied along the $a^*$ axis, and the field was applied along the $c$ axis.  The excitation current was $5\,$mA and the measurement frequency was $10-20\,$Hz. Because the polarity of the magnetic field cannot be reversed in this setup, we removed the sample from the cryostat and manually rotated the stage by $180\,$ degrees before re-inserting once again. 

The high-field data was compared to measurements on the same sample, with the same wiring, acquired in our in-house PPMS cryostat. We noticed offsets of the Hall resistivity ($\sim10\,\%$) and longitudinal resistivity (up to $\sim 25\,\%$) which likely occurred due to phase rotations in the imperfect electrical circuit used for the high-field experiment. We corrected these offsets by scaling the high-field results to the in-house data (Fig. \ref{fig:highfield}). We emphasize that the low-field measurements presented in the main text did \textit{not} suffer from similar problems with the electric circuit.

Magnetoresistance $\rho_{xx}$ and Hall conductivity $\sigma_{xy}$ are shown in Fig.~\ref{fig:highfield}a and b, respectively. The high field data (solid colors) are compared to in-house low-field measurements (shaded colors). The change of $\rho_{xx}(H)$ is relatively weak and quasi-linear in the high-field regime. Meanwhile, $\sigma_{xy}$ bends strongly at the lowest $T$, and curves more gently at $T=10$ and $20\,$K. This result is generally in agreement with the expectation for the geometrical Hall effect, where $\sigma_{xy}^A$ constitutes an additional contribution to $\sigma_{xy}$, which appears only below the ordering transition ($\sim 5\,$K) and is suppressed at very large $H$. The data also confirm the result of previous magnetization measurements \cite{gorbunov_electronic_2014}, which indicates that there are no additional field-induced transitions above $\mu_0H = 1.5\,$T in this compound when the field is applied along the $c$ axis.

A quantitative analysis of the high-field data remains challenging, in large part due to the rather complex behavior of the normal Hall effect at $\mu_0 H>9\,$T. This is related to the presence of a small carrier pocket with moderately high carrier mobility ($\mu \sim 300-800\,\text{cm}^2/(\text{Vs}$), which leads to bending of the $\sigma_{xy}$ curves already at $T\ge 10\,$K (Fig.~\ref{fig:highfield}b). We modeled the curves at $T\ge 10\,$K using the two-band Drude model for the normal Hall conductivity $\sigma_{xy}^N = aB+bB/\left(1+\left(\mu B\right)^2\right)$ (Ref.~\cite{xiong_high_2012}). Here, $B = \mu_0H$. From our data, it cannot be excluded that $\mu$ increases in magnitude below the transition to long-range order ($\sim 5\,$K). The strong bending of $\sigma_{xy}(T = 2.5\,\text{K}, H)$ may be at least partially due to $\sigma_{xy}^N$. 

The presence of the small Fermi surface pocket was also confirmed using Shubnikov-de Haas (SdH) quantum oscillation experiments (Fig. \ref{fig:highfield}c). The oscillations were not resolved in the $\rho_{xx}$ channel. We proceeded by subtracting a fifth-order polynomial background from the Hall resistivity to obtain the oscillatory part $\rho_{yx}^{\textrm{osc}}$. This quantity was then subjected to a smoothing algorithm to improve the signal-to-noise ratio (SNR). Finally, we calculated $\sigma_{xy}^{\textrm{osc}} = \rho_{yx}^{\textrm{osc}}/\left(\rho_{xx}^2 + \rho_{yx}^2\right)$. This approach is only valid in the limit $\rho_{yx}\ll \rho_{xx}$.

The oscillatory part of the Hall conductivity was fitted using the standard Landau-Lifshitz-Kosevich (LLK) expression, which is typically applied in the case of $\sigma_{xx}$ (Ref.~\cite{roth_semiconductors_1966, shoenberg_magnetic_1984}),
\begin{equation}
\frac{\sigma_{xy}^{\textrm{osc}}}{\sigma_{xy}} = \left(\frac{\hbar \omega_c}{2\epsilon_F}\right)^{1/2}\frac{\lambda}{\sinh\lambda}\exp\left(-\lambda_D\right)\,\cos\left[\frac{2\pi \epsilon_F}{\hbar\omega_c}+\varphi\right]\ \mathrm{,}
\end{equation}
with $\lambda = 2\pi^2 k_B T/\hbar\omega_c$ and $\lambda_D = 2\pi^2k_BT_D/\hbar\omega_c$. The cyclotron frequency and the Dingle temperature are given by $\omega_c = eB/m^{*}$ and $T_D = \hbar/\left(2\pi k_B\tau\right)$, respectively.  The carrier relaxation time $\tau$ is related to the mobility viz. $\mu_\text{SdH} = e\tau/m^{*}$.

From the LLK fit, we extract the Fermi surface cross-section $S_F = 2\pi \epsilon_Fm^{*}/(\hbar e) = 83\,$T and $\mu_\text{SdH} \approx 900\pm 400\,\text{cm}^2/(\text{Vs})$. The error bar of $\mu_\text{SdH}$ is sizable due to the rather low SNR and the small size of the oscillatory amplitude. Finally, the evolution of the SdH oscillations with temperature argues against a major modification of the electronic structure at the onset of long-range magnetic order ($T\sim 5\,$K).

\end{document}